\documentclass[11pt,twoside]{article}
\usepackage{asp2010}
\usepackage{graphicx}

\resetcounters

\begin{document}

\title{Data reduction pipeline for the MMT Magellan Infrared Spectrograph
}
\author{Igor~Chilingarian, 
Warren~Brown, Daniel~Fabricant, 
Brian~McLeod, John~Roll,
and Andrew~Szentgyorgyi
\affil{Smithsonian Astrophysical Observatory, 60 Garden St.,
Cambridge, MA 02138, USA}}

\begin{abstract} We describe principal components of the new spectroscopic
data pipeline for the multi-object MMT/Magellan Infrared Spectrograph
(MMIRS).  The pipeline is implemented in {\sc idl} and {\sc C++}.  The
performance of the data processing algorithms is sufficient to reduce a
single dataset in 2--3~min on a modern PC workstation so that one can use
the pipeline as a quick-look tool during observations.  We provide an
example of the spectral data processed by our pipeline and demonstrate that
the sky subtraction quality gets close to the limits set by the Poisson
photon statistics.
\end{abstract}

\section{Introduction} 

MMIRS \citep{McLeod+12} is a new near-infrared spectrograph and imager which
can operate at f/5 focii of either the MMT or the Magellan Clay 6.5m
telescopes.  MMIRS addresses a 4$\times$7~arcmin field of view for
multi-object spectroscopy (MOS) and is equipped with a 2K$\times$2K HAWAII-2
array.  It can also operate as a single-object spectrograph with a 7~arcmin
long 0.2--3~arcsec wide slit.  Here we present the data reduction pipeline
which handles data obtained in the MOS and long slit modes.

\section{Pipeline design and implementation}

\begin{figure}
\includegraphics[angle=-90,width=\hsize]{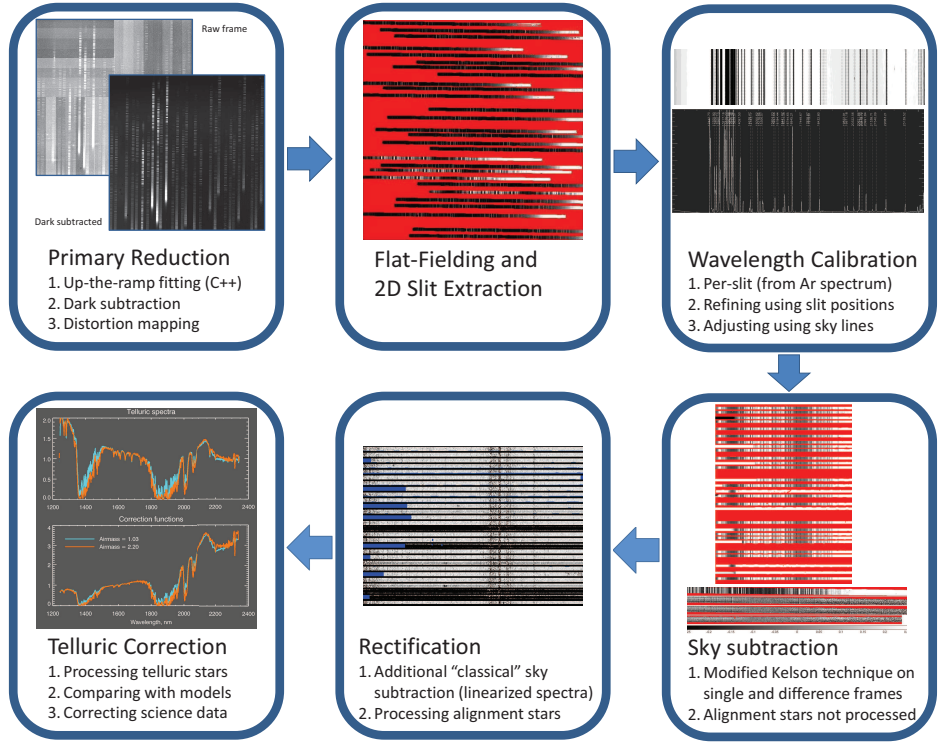}
\caption{Pipeline steps executed for every observing block.\label{figP58_1}}
\end{figure}

The MMIRS spectroscopic data reduction pipeline is a stand-alone {\sc idl}
package with the most computationally intensive step implemented in {\sc
C++}.  It is controlled by a task control file having a format resembling a
FITS header, i.e.  a set of keyword--value pairs.  This file can be created
either manually in a text editor or automatically from a database containing
observational metadata generated after each observing run.  Then it can be
edited in order to perform specific data reduction steps or adjust certain
parameters of data processing algorithms.  The following steps are
executed for every observing block represented by a pair of spectra
spatially dithered along the slit (see Fig.~\ref{figP58_1}).

 \emph{Fitting the up-the-ramp slope} in every pixel and obtaining a
2-dimensional image. At present, this step is performed by the {\sc
mmfixen} package implemented in {\sc C++}. It allows us to detect cosmic ray
hits, recover the counts in most cases unless the cosmic ray hits the
detector in the very beginning of exposure, as well as to handle saturation
and detector non-linearity at high counts.

 \emph{Dark subtraction and scattered light subtraction} is done using
up-the-ramp processed 2D frames.  We average several dark frames which are
obtained during observations for every read-out setup and subtract them from
science and calibration frames.  Then we create subtracted dithered pairs of
images which are then processed through the pipeline along with individual
exposures.  To subtract the scattered light (only for MOS data), the
pipeline analyses the MOS mask definition files, then uses them to trace the
spectra and gaps between them on a dark subtracted spectral flat field
exposure.  Then, a smooth background model is constructed from gap counts in
all types of frames (science, calibrations, telluric standards) using
low-order polynomials.

 \emph{Mapping the optical distortions} is done using spectral traces
from the previous step.  We use a low order 2D polynomial approximation to
describe distortions coming from both, collimator and camera. The distortion
map accounts not only for optical distortions but also for the imperfect
positioning of the grism causing the tilt of spectra with respect to the
detector edge by up-to 0.7~deg).

 \emph{Extraction of 2D spectra and flat fielding} is done by analysing mask
definition files and extracting the bounding boxes of every spectrum.  Then
we create a normalized flat field and correct all exposures by it.  This is
a critical step in the NIR data reduction as pixel-to-pixel variations
in the HAWAII-2 detector reach 35~per~cent.

 \emph{The wavelength calibration} is done in several steps. At first,
the argon lines from the comparison spectrum frame are automatically
identified and the initial wavelength solution is approximated using a low
order 2D polynomial.  This step applied to the entire frame in the long-slit
mode or to every extracted slit in the MOS mode.  Alignment stars where
spectral lines are too broad are not processed at this stage.  The second
step of the wavelength calibration applies only to MOS data: we approximate
the coefficients in all slits simultaneously as low order 2D polynomials of
the (X,Y) position on the mask.  This allows us to handle alignment star
boxes by evaluating the polynomial at their measured positions.  The last
step which is also applied to the long-slit data accounts for imperfect
cutting of the slit masks: we create a ``template spectrum'' using the
science frame assuming that targets are faint and airglow lines are
prominent and bright (or arc lines in case of bright targets or short
exposure times) which is then cross-correlated at every position across
dispersion in order to determine small residual shifts which are then
included into the final wavelength solution.

 \emph{Sky subtraction} is one of the most critical steps in the NIR data
reduction.  NIR night sky spectrum contains rapidly changing very bright
emission lines (mostly hydroxyl, OH) as well as the continuum background. 
Targets observed with MMIRS with integration times of many hours are up-to
hundreds of times fainter than the night sky level.  Therefore high quality
sky subtraction is essential.  We use a hybrid approach using a combination
of the classical ``difference imaging'' and a recently developed sky
subtraction technique.  Difference image contains the residual night sky
emission originating from the temporal variation of OH line fluxes which may
reach tens of per~cent over 5--10~min.  We apply the modified
\citet{Kelson03} technique to the difference images in order to remove the
residual night sky emission.  This technique uses non-resampled spectra and
precise pixel-to-wavelength mapping in order to create the oversampled model
of the night sky spectrum which is then parametrized using $b$-splines and
evaluated at every position at every slit.  It allows one to avoid artefacts
originating from the interpolation of sharp undersampled OH line profiles. 
However, this requires slit images on the detector frame to have sufficient
curvature in order to well sample the line contours.  While it works for
long-slit observations, in the MOS mode the situation is different.  For a
typical slit length of 7--8~arcsec, the curvature in the central slit images
is close to zero and it is between 1 and 2 pixels in the edges of the field
of view.  Therefore, in the MOS mode we use all the slits simultaneously
assuming that their widths are the same and create the sky model using a 3D
$b$-spline/polynomial parametrization where the $b$-splines describe the
night sky model along the wavelength while across the field of view we use a
2D Legendre polynomial as a function of X and Y slit positions on the mask
to account for smooth variations of $b$-spline coefficients.  By using this
approach we get the sky subtraction quality close to the limit set by the
Poisson photon statistics.

\emph{Final cosmic ray cleaning, linearisation and rectification} is
performed on sky subtracted frames.  We use the modified Laplacian filtering
technique \citep{vanDokkum01} which handles both, negative and positive
``hits'' on difference images.  At the end, we put the spectra into the
linear wavelength scale and rectify the spectra by correcting the optical
distortions which we mapped earlier.

\emph{Telluric absorption correction} is the last step of the pipeline
reduction where we use observations of a hot (typically A0V) star and
compare it against the stellar atmosphere model in order to measure the water
vapor absorption and correct science spectra for it. It also automatically
calibrates the flux along the wavelength, although no absolute flux
calibration is done.

The data post-processing is run after the pipeline processing and it
includes the following steps: (1) co-adding multiple observing blocks; (2)
extraction of 1D spectra.  The final data product is available in several
formats: (a) multi-extension FITS file with one extension per 2D extracted
calibrated co-added spectrum; (b) single-extension FITS file with all
extracted 1D spectra; (c,d) a Euro3D-FITS file for 2D and 1D extracted
spectra.  The first two files are accompanied with the files in the same
format providing flux uncertainties, while this information is stored in the
Euro3D-FITS format.  MMIRS MOS spectra in the Euro3D-FITS format contain
metadata making them Virtual Observatory compliant
\citep{CBLM06,Chilingarian+08b}.

\section{Example of a fully reduced dataset}

\begin{figure}
\includegraphics[angle=-90,width=\hsize]{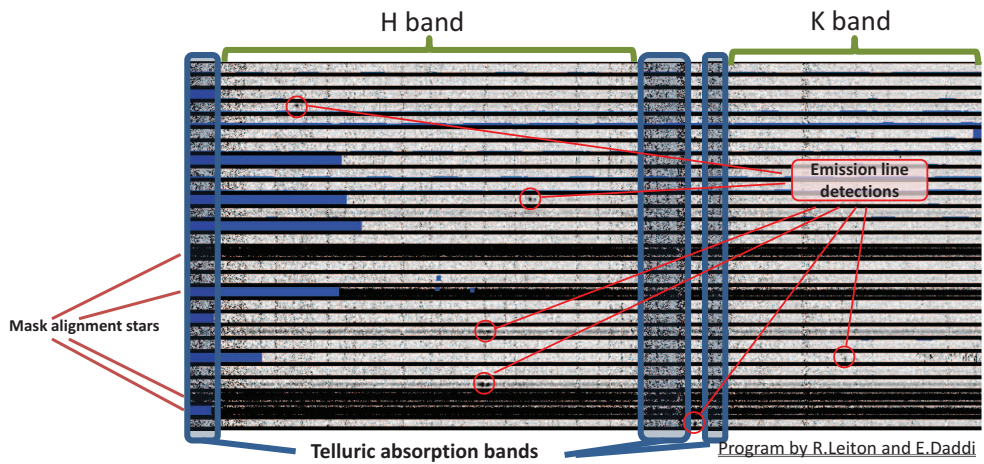}
\caption{Deep spectroscopy of a field containing high-redshift (z =
1.5--2.0) active galactic nuclei in the magnitude range of $22<m_H<20$~mag\label{figP58_2}}
\end{figure}

In Fig.~\ref{figP58_2} we present an example of a dataset obtained in
Dec~2011 in the low-resolution HK mode at the 6.5~m Magellan telescope. The
integration time was 7h~20min. Our excellent sky subtraction quality
leads to detections of faint galaxies ($m_H>20$~mag).

\bibliography{bibP58}

\begin{thebibliography}{}
\expandafter\ifx\csname natexlab\endcsname\relax\def\natexlab#1{#1}\fi
\expandafter\ifx\csname url\endcsname\relax
  \def\url#1{\texttt{#1}}\fi
\expandafter\ifx\csname urlprefix\endcsname\relax\def\urlprefix{URL }\fi
\providecommand{\eprint}[2][]{\url{#2}}

\bibitem[{{Chilingarian} et~al.(2006){Chilingarian}, {Bonnarel}, {Louys}, \&
  {McDowell}}]{CBLM06}
{Chilingarian}, I., {Bonnarel}, F., {Louys}, M., \& {McDowell}, J. 2006, in
  Astronomical Data Analysis Software and Systems XV, edited by C.~{Gabriel},
  C.~{Arviset}, D.~{Ponz}, \& S.~{Enrique}, vol. 351 of Astronomical Society of
  the Pacific Conference Series, 371

\bibitem[{{Chilingarian} et~al.(2008){Chilingarian}, {Bonnarel}, {Louys},
  {Zolotukhin}, {Royer}, {Jegouzo}, {Le Sidaner}, {Fernique}, \&
  {Boch}}]{Chilingarian+08b}
{Chilingarian}, I., {Bonnarel}, F., {Louys}, M., {Zolotukhin}, I., {Royer}, F.,
  {Jegouzo}, I., {Le Sidaner}, P., {Fernique}, P., \& {Boch}, T. 2008, in
  Astronomical Spectroscopy and Virtual Observatory, edited by {M.~Guainazzi \&
  P.~Osuna}, 125. \eprint{arXiv:0711.0412}

\bibitem[{{Kelson}(2003)}]{Kelson03}
{Kelson}, D.~D. 2003, \pasp, 115, 688. \eprint{arXiv:astro-ph/0303507}

\bibitem[{{McLeod} et~al.(2012){McLeod}, {Fabricant}, {Nystrom}, {McCracken},
  {Amato}, {Bergner}, {Brown}, {Burke}, {Chilingarian}, {Conroy}, {Curley},
  {Furesz}, {Geary}, {Hertz}, {Martini}, {Matthews}, {Norton}, {Park}, \&
  {Roll}}]{McLeod+12}
{McLeod}, B., {Fabricant}, D., {Nystrom}, G., {McCracken}, K., {Amato}, S.,
  {Bergner}, H., {Brown}, W., {Burke}, M., {Chilingarian}, I., {Conroy}, M.,
  {Curley}, D., {Furesz}, G., {Geary}, J., {Hertz}, E., {Martini}, P.,
  {Matthews}, A., {Norton}, T., {Park}, S., \& {Roll}, J. 2012, accepted to
  \pasp. \eprint{arXiv:1211.6174}

\bibitem[{{van Dokkum}(2001)}]{vanDokkum01}
{van Dokkum}, P.~G. 2001, \pasp, 113, 1420. \eprint{arXiv:astro-ph/0108003}

\end{thebibliography}

\end{document}